\documentclass[aps,prl,twocolumn,superscriptaddress,nofootinbib]{revtex4-1}
\usepackage{amsmath}
\usepackage{graphicx}
\usepackage{amsbsy}
\usepackage{amssymb}
\usepackage{palatino,avant,graphicx,color}
\usepackage[none]{hyphenat}
\usepackage{tikz}
\usepackage{xy}
\usepackage{comment}
\usepackage[none]{changes}
\usepackage{wrapfig}
\usepackage{float}
\usepackage{mciteplus}
\usepackage{tabularx}
\usepackage{natbib}
\usepackage{hyperref}
\usepackage{mathtools}
\usepackage[artemisia]{textgreek}
\usepackage{newunicodechar}

\newcommand{\cmmnt}[1]{}

\begin{document}

\title{Nonlocal and Nonlinear Surface Plasmon Polaritons and Optical Spatial Solitons induced by the Thermocapillary Effect}

\author{Shimon Rubin}
\email{rubin.shim@gmail.com}  
\author{Yeshaiahu Fainman}
\affiliation{Department of Electrical and Computer Engineering, University of California, San Diego, 9500 Gilman Dr., La Jolla, California 92023, USA}

\begin{abstract}

We study the propagation of surface plasmon polaritons (SPPs) on a metal surface which hosts a thin film of a liquid dielectric. 
The Ohmic losses that are inherently present due to the coupling of SPPs to conductors' electron plasma,
induce temperature gradients and fluid deformation driven by the thermocapillary effect, which lead to a nonlinear and nonlocal change of the effective
dielectric constant. The latter extends beyond the regions of highest optical intensity and constitutes a novel thermally self-induced mechanism that affects the propagation of the SPPs. We derive the nonlinear and nonlocal Schr{\"o}dinger equation (NNLSE) that describes propagation of low intensity SPP beams,  
and show analytically and numerically that it supports a novel optical spatial soliton excitation.

\end{abstract}

\maketitle

\textbf{Introduction:} Surface plasmon polaritons (SPPs) are electromagnetic excitations that propagate at the interface between a metal and a dielectric material \cite{raether2013surface,maier2007plasmonics}.
The unique properties of the SPPs that enable one to concentrate light in a subwavelength region around the interface and SPPs' sensitivity to changes of the dielectric constant, have motivated numerous theoretical and experimental studies over the last few decades with a broad range of applications in biosensing \cite{anker2008biosensing}, medicine \cite{gobin2007near}, thermal and photoimaging \cite{mecklenburg2015nanoscale,goykhman2011locally}, and solar energy \cite{atwater2010plasmonics,linic2011plasmonic}.
The inherent Joule heat generation due to Ohmic losses 
leads to an increase of the metal's temperature and affects the properties of nearby objects by heat conduction. 
While previous works investigated heating effects on fluids due to SPP heat generation, such as generation of Rayleigh-B{\'e}nard convection, \cite{roxworthy2014understanding,donner2011plasmon}, thermophoretic migration of suspended particles \cite{ndukaife2016long,lin2017thermophoretic},  gas-fluid phase transition \cite{govorov2007generating}, generation of microstructures in polymer films \cite{rontzsch2007thin} and transport of liquid droplets \cite{passian2006nonradiative} - to the best of our knowledge the effect of the fluid on SPP due to self-induced heating hasn't been reported to date.
\begin{figure}[th]
	\includegraphics[scale=0.19]{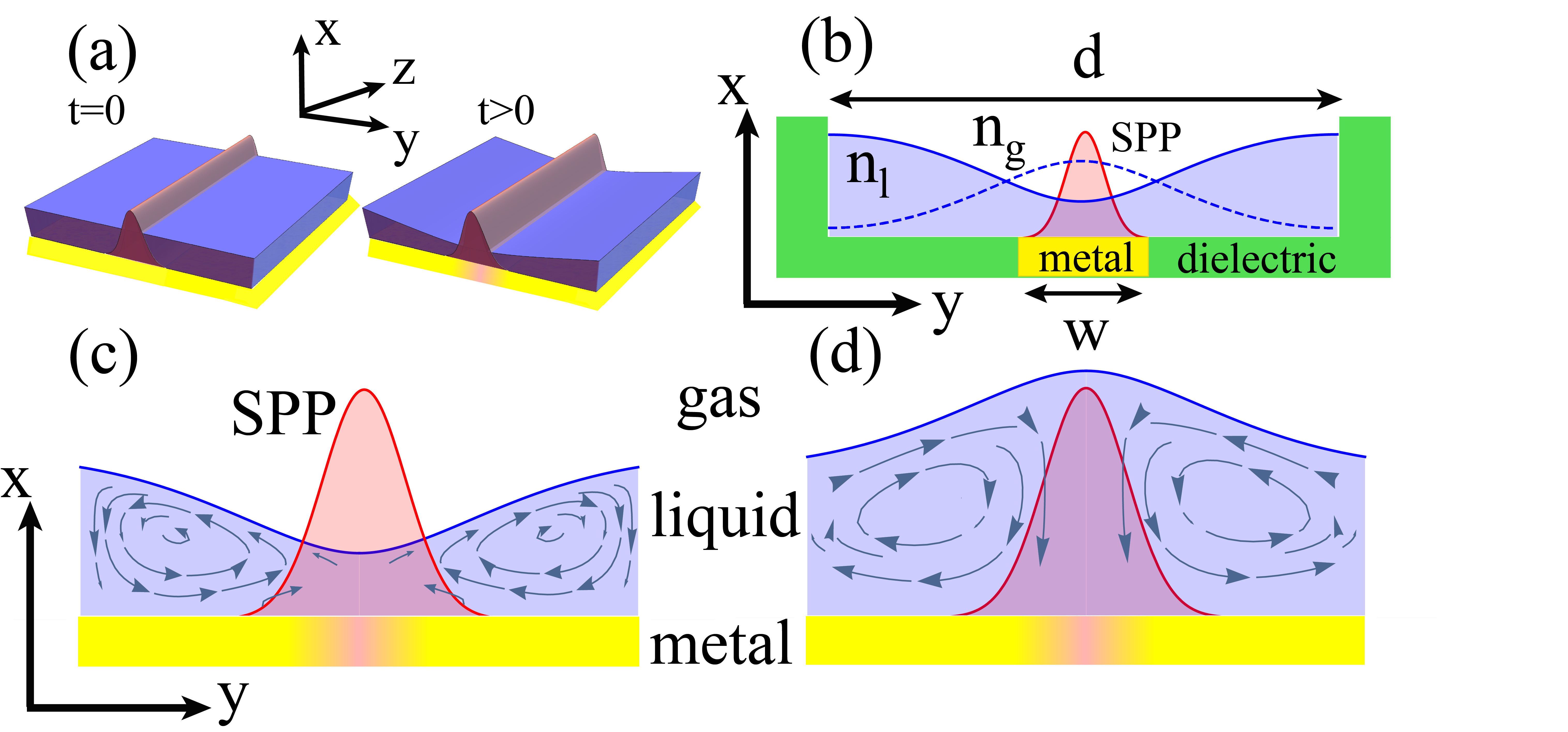}
    \caption{(a) 3D scheme of a propagating SPP along the $z$ axis on a metal surface near a thin liquid film; left - uniform thin film of thickness $h_{0}$ prior to heating effects; right - deformed thin film after SPP generates nonuniform temperature field. (b) Normal cross section of a propagating SPP along a metal of a finite width $w$ embedded within a fluidic slot of width $d$, (c,d) Normal sections presenting thermocapillary flows (blue arrows) and the deformed free interface for the positive and negative Marangoni constant, respectively, for the case $w$ and $d$ are larger than the width of the SPP beam.     
}
    \label{Setup}
\end{figure}

In this work we theoretically study the interaction between a propagating SPP along a planar metal surface and an adjacent thin film of a dielectric liquid. 
Figure \ref{Setup}(a) presents a schematic description of the problem; a spatially nonuniform SPP beam propagates along the $z$ direction and locally heats the metal, which in turn heats the gas-fluid interface. Local increase of the free interface temperature leads to a surface tension gradient
and triggers the thermocapillary effect \cite{pearson1958on} (a special case of the Marangoni effect \cite{marangoni1871}), manifested by thermocapillary flows with a distinctive shape of B{\'e}nard cells \cite{bernard1900tourbillons} and deformation of the free interface. 
A local temperature increase usually leads to a decrease of the surface tension in the hotter region, and to interfacial flows from the hotter region to the colder region, though few liquids
are known to exhibit an opposite behavior of the surface tension \cite{cheng2017surface}, both illustrated in Figs.\ref{Setup}(c) and \ref{Setup}(d).
In case the liquid film is thinner than the penetration depth of the SPP into the bulk, the liquid deformation is coupled back to the Maxwell equations through the changes of the liquid's dielectric constant, and together with the heat transport form a complete set of coupled equations. 
This novel SPP-fluid coupling mechanism induces changes in the geometrical shape of the thin film of a liquid dielectric, which is is fundamentally different from the traditional thermo-optical effect, where the source of dielectric function changes stems from changes of material density and polarization.
Importantly, the resultant change of the liquid's refractive index is spatially nonlocal, in a sense that the induced fluid deformation extends beyond
the regions of highest optical intensity to more distant regions. Prominent mechanisms that are known to admit a light-induced nonlocal index response are:
charge transport in photorefractive crystals \cite{duree1993observation}, atomic diffusion in atomic vapors \cite{suter1993stabilization,skupin2007nonlocal}, heat transport and changes of volume or atomic polarizability \cite{rotschild2005solitons}, 
molecular long-range interaction in nematic liquid crystals \cite{mclaughlin1995paraxial, conti2003route, assanto2003spatial}, coupling of lasing disordered resonators via directional stimulated emission \cite{leonetti2013non} and quantum effects on the Thomas-Fermi screening length \cite{marinica2012quantum,zuloaga2009quantum}. Several of these nonlocal mechanisms support optical solitons, 
which are localized intensities of light intensity due to a balance between
diffraction and nonlinearity of the medium, and have been in the focus of an active research for the last few decades (see Refs. \cite{chen2012optical,kivshar2003optical} and references within). 

In this Letter we derive NNLSE for an SPP under thermocapillary induced nonlocal nonlinearity in the dielectric material and then
show that it supports a novel spatial soliton excitation - an SPP that propagates on a metal covered with a thin liquid film 
and a dynamical fluidic plasmonic photonic crystal (FPPC) with intensity tunable band gap. We show that our model naturally admits a strongly nonlocal limit of the Snyder-Mitchell accessible soliton model \cite{snyder1997accessible}, and present numerical results that describe diffraction of two SPP beams analogous to the Young double slit experiment.
 
\textbf{Governing equations for the non-linear media:}
Our starting point is the Navier-Stokes equations for a noncompressible fluid of viscosity $\mu$, mass density $\rho$, velocity field $u_{i}$ and stress tensor $\tau_{ij}$ for Newtonian fluid given by \cite{landau1987fluid}
\begin{equation}
	\rho \left( \partial_{t}u_{i}+ u_{j} \partial_{j} u_{i} \right) = \partial_{j} \tau_{ij}; \quad i,j=x,y,z,
\label{NavierStokes}    
\end{equation}
%
The free surface of the thin film, which rests on a metal surface [Fig.\ref{Setup}(a)], satisfies the following stress balance equation \cite{landau1987fluid}
\begin{equation}
	\tau_{ij} n_{j}  = \sigma n_{i} \vec{\nabla} \cdot \hat{n} - \vec{\nabla}_{\parallel} \sigma,
\label{MatchingConditions}    
\end{equation}
where, $\sigma$ is the surface tension, $\vec{\nabla} \cdot \hat{n}$ is the divergence of the normal 
and $\vec{\nabla}_{\parallel}$ stands for a gradient with respect to the in-plane coordinates ($y,z$). In this work we assume that the surface tension depends on temperature via \cite{levich1962physicochemical}
\begin{equation}
	\sigma(T)=\sigma_{0}-\sigma_{T} \Delta T; \quad \Delta T \equiv T-T_{0},
\label{SurfaceTensionGrad}
\end{equation}
where $\sigma_{T}$ is the Marangoni constant known to exhibit nearly temperature independent values over a large range of temperature for many materials \cite{adamson1990physical}, 
and the temperature field in the metal, $T^{m}$, is governed by the following 2D equation
\begin{equation}
	\dfrac{\partial T^{m}}{\partial t} - D_{th}^{m} \nabla^{2}_{\parallel} T^{m} = \dfrac{\Delta T}{I_{0} \tau_{th}}  \chi  I; \quad \chi \equiv \dfrac{\alpha_{th}^{m}  d^{2} I_{0}}{k_{th}^{m} \Delta T},
\label{HeatEquationDiffusion2c}    
\end{equation}
where the superscript $m$ stands for quantities in the metal \textcolor{black}{and $I$ is an optical intensity of typical strength $I_{0}$}. 
Here, $D_{th}^{m}=k_{th}^{m}/(\rho^{m} c_{p}^{m})$ is the heat diffusion coefficient;  $\rho^{m}$, $c_{p}^{m}$, $k_{th}^{m}$, $\alpha_{th}^{m}$ are the mass density, specific heat,
heat conductance, and optical absorption coefficient, respectively; $\tau_{th}=d^{2}/D_{th}^{m}$ is the typical time scale; $d$ is the typical length scale along the in-plane direction; $\chi$ is the dimensionless intensity of the heat source.

Applying \textcolor{black}{low Reynolds number and thin film assumptions (i.e., lubrication approximation) \cite{happel2012low,howison2005practical}, 
allows us to neglect the inertial terms in the Navier-Stokes equations, Eq.(\ref{NavierStokes}), and drop the in-plane derivatives
relative to the normal derivative. Together with the thin film limit
of the matching conditions, Eq.(\ref{MatchingConditions}) \cite{oron1997long}, yields the following equation for the thin film deformation $\eta$},
\begin{equation}
	\dfrac{\partial \eta}{\partial t} + D_{\sigma} \nabla^{4}_{\parallel}  \eta= - \dfrac{\sigma_{T} h_{0}^{2}}{2 \mu}  \nabla^{2}_{\parallel} T^{m}; \quad D_{\sigma} \equiv \sigma_{0} h_{0}^{3}/(3 \mu),
\label{ThinFilmEqU33}
\end{equation} 
which includes the effects of surface tension and thermocapillarity.
Effects of gravity are negligible on a microscale and expected to emerge on a much larger scales comparable to the capillary length, \cite{de2013capillarity}, whereas nonretarded van der Waals interaction can be neglected for films with thickness above $100 \text{ nm}$ \cite{israelachvili2011intermolecular}.
The forces on a dielectric film due to nonhomogeneity of the dielectric function on the free surface and electrostriction \cite{landau2013electrodynamics}, are expected to have much lower magnitude than the thermocapillary effect \cite{SuppInfo}.

Taking advantage of the linearity of the thermal transport and the thin film equations and assuming quasistatic temperature field distribution,
which for transient problems typically holds at $t>\tau_{th}$, we can represent the deformation $\eta$ 
in terms of the Green's function $G_{l}$ of Eq.(\ref{ThinFilmEqU33}) and intensity as \cite{SuppInfo} 
\begin{equation}
	\eta (\vec{r}_{\parallel},t)/h_{0} = - \text{M}
    \int d\vec{r}^{\prime}_{\parallel} dt^{\prime} \dfrac{1}{\tau_{th}} 
    G_{l}(\vec{r}_{\parallel}-\vec{r}^{\prime}_{\parallel},t-t^{\prime}) I(\vec{r}^{\prime}_{\parallel},t^{\prime})/I_{0}.
\label{GreenGreen22}    
\end{equation}
Here, $\text{M} \equiv \text{Ma} \cdot \text{\textchi}/2$ and $\text{Ma}=\sigma_{T}\Delta T h_{0}/(\mu D_{th}^{m})$ is the dimensionless Marangoni number which represents the ratio between the surface tension stresses due to the thermocapillary effect, and dissipative forces due to fluid viscosity and thermal diffusivity. 
The typical values of the time scales $\tau_{l}$, $\tau_{th}$ and $\tau_{el}$, that govern the transport of liquid, heat and propagation of SPP, respectively, satisfy 
\begin{equation}
	\tau_{el} \ll \tau_{th} \ll \tau_{l} = d^{4}/ D_{\sigma}.
\label{hierarchy}    
\end{equation}  
Indeed, for the following values $d=1$ \textmu m, $h_{0}=0.25$ \textmu m, 
$D_{th}^{m}=10^{-4}$ m$^{2}$s$^{-1}$, $D_{th}^{ }=10^{-7}$ m$^{2}$s$^{-1}$, $\mu=10^{-3}$ Pa$\cdot$s, $\sigma_{T}=10^{-4}$ Nm$^{-1}$K$^{-1}$, $\sigma_{0} = 10^{-3}$ N\text{ }m$^{-1}$
we learn that $\tau_{l} = 10^{-4}$ s, 
$\tau_{th} = 10^{-8}$ s, are much larger than 
$\tau_{el} = 1/ \omega = 10^{-14} \text{ s}$ and Eq.(\ref{hierarchy}) holds. The additional time scale that governs heat diffusion from the metal to the free surface, $h_{0}^{2}/D_{th}$, is on the order of magnitude $10^{-6} \text{ s}$, which is still much smaller than $\tau_{l}$. 

\textbf{Nonlocal and nonlinear SPP:} We now utilize perturbation theory accounting for the propagation and diffraction of an SPP on a metal-dielectric interface, incorporating,
dissipation as well as nonlinear and nonlocal effects.
We start our analysis from the time independent Maxwell equations for TM waves.
Combining the sourceless Maxwell equations, yields the following equations for the electric field, $E_{i}^{m,d}$ \cite{raether2013surface,maier2007plasmonics}, 
\begin{equation}
	\partial_{ij}^{2}E_{j}^{(m,d)}-\partial_{jj}^{2}E_{i}^{(m,d)}=k_{0}^{2} \epsilon(\vec{r}) E_{i}^{(m,d)}; \quad k_{0}=\omega/c,
\label{MaxwellCombinedCartesian}    
\end{equation}
where $i$ labels the different equations ($i=x,y,z$), $j$ is a summation index which runs on all values $j \neq i$, and $m,d$ stand for the metal and dielectric regions, respectively. 
Employing the depth averaged approximation \cite{campbell1998quantitative}, \textcolor{black}{we treat the gas-fluid bilayer as a single media with an effective index calculated by averaging the index above the metal ($x>0$) weighted by the decay factor $2q_{d} e^{-2 q_{d}x}$. Specifically, integrating the index distribution $n_{l}$ (liquid) between $0 < x < h_{0}+\eta$ and $n_{g}$ (gas) for $x > h_{0}+\eta$ yields the corresponding changes of the depth averaged index and the dielectric constant \cite{SuppInfo}} 
\begin{equation}
\begin{split}
	 &\Delta n_{D}(\eta) = \tilde{b} \eta(\vec{r}_{\parallel},t) /h_{0} ; \text{ } \tilde{b}= 2 q_{d} h_{0} (n_{l} - n_{g} ) e^{-2q_{d}h_{0}};
\\
	& \Delta \epsilon_{D}(\eta) = b \eta(\vec{r}_{\parallel},t) /h_{0} ; \text{ } b= 2n_{0} \tilde{b},
\end{split}     
\label{epsilonD}
\end{equation}
respectively. Here, we kept the leading term in the $\eta/h_{0}$ series, $q_{d}^{2} = \beta_{0}^{2}(1-\epsilon_{D})$, $\beta_{0}=k_{0}\sqrt{\epsilon_{m}\epsilon_{D}/(\epsilon_{m}+\epsilon_{D})}$, $n_{0} = \sqrt{\epsilon_{D}}= n_{l} - (n_{l}-n_{g}) e^{-2q_{d}h_{0}}$ and the dimensionless deformation, $\eta/h_{0}$, is determined by Eq.(\ref{GreenGreen22}). 
Employing perturbative expansion in dimensionless number $\text{M}$ of the governing equations for SPP in the metal, dielectric with depth averaged dielectric function, $\Delta \epsilon_{D}$, and the matching conditions between metal and dielectric, we derive the following NNLSE \cite{SuppInfo}
\begin{equation}
	2i \beta_{0} \dfrac{\partial A}{\partial z} + \dfrac{\partial^{2} A}{\partial y^{2}} +  \tilde{\chi}_{TC}  A \int d\vec{r}^{\prime}_{\parallel} dt^{\prime} G_{l}(\vec{r}_{\parallel} - \vec{r}^{\prime}_{\parallel},t-t^{\prime}) \vert A \vert^{2}  =0,
\label{Schrodinger2}    
\end{equation}
where $\tilde{\chi}_{TC}= k_{0}^{2} \chi_{TC}/(I_{0} \tau_{h})$ and $A(y,z)$ is the envelope of the SPP beam \cite{SuppInfo}.
Here, $\chi_{TC}$ is a dimensionless number, given by $\chi_{TC} = f b \text{M}$
and incorporates the effects of thermocapillarity, kinematics of the index averaged model and plasmonic enhancement \cite{marini2010ginzburg} through the dimensionless numbers $M$, $b$ and $f$, respectively (see Ref. \cite{SuppInfo} for expression for $f$ and Refs. \cite{feigenbaum2007plasmon, davoyan2009self} for an alternative derivation).

\textbf{The limit of local interaction:}  
Consider the case schematically presented in Fig.\ref{Setup}(b), where SPP of vacuum wavelength $\lambda$ is restricted to propagate along a metal slab of width $w$, which is smaller than the width of the fluidic slot $d$. Furthermore, we assume that $w \gg \lambda$, which allows us to neglect edge and other effects due to strong lateral confinement which lead to an enriched mode spectrum \cite{berini2009long}. Therefore, we can assume that SPP admits the form of a nondiffracting beam, and following \cite{mihalache1987exact} the corresponding matching conditions at $x=0$ lead to the following dispersion relation 
\begin{equation}
	\beta_{0} \epsilon_{m} E_{z}^{(m)} \Big \vert_{x=0} = i q_{m} \left( n_{0}^{2} + \Delta \epsilon_{D}(\eta) \right) E_{x}^{(d)} \Big \vert_{x=0}.
\label{MatchingCondCombined}    
\end{equation}
Next, let us determine the temperature distribution and the resulting thin film deformation due to an SPP that begins to propagate along the slab at $t=0$. To this end, we determine the Green's functions, $G_{th}$ and $G_{l}$ which satisfy, respectively, Eqs.(\ref{HeatEquationDiffusion2c}) and (\ref{ThinFilmEqU33}) with a source term $\delta(x-x_{0}) H(t)$, where $H(t)$ is Heaviside function. For convenience, we consider the case where the thin film forms an angle $\pi/2$ with the walls at $x=0,d$, 
whereas the temperature field satisfies Dirichlet boundary conditions at the edges $x=(d \pm w)/2$. Employing the closure relation \cite{arfken1999mathematical}, we derive the corresponding expressions for $G_{l}$ and $G_{th}$ \cite{SuppInfo}, which upon convolving with Gaussian intensity leads to the following closed form expression for $\eta$ \cite{SuppInfo}
\begin{equation}
	\dfrac{\eta(y,t)}{h_{0}} = - \frac{2 d \text{M} \tau_{l}I}{3\pi^{6} \tau_{th}I_{0}}  \sum\limits_{n=1}^{\infty} \frac{(-1)^{n}}{\lambda_{n} n^{4}}
   \varphi_{n}(y)  \varphi_{n}(\tfrac{w}{2})
    \left( 1 - e^{-\lambda_{n} \frac{t}{\tau_{l}}} \right).
\label{SolThinFilm}
\end{equation}
Here, $\lambda_{n}$ is a constant (see Ref. \cite{SuppInfo}) and $\varphi_{n}(y)=\sqrt{2/d} \cos(n \pi y/d)$ is the set of the eigenfunctions associated with the corresponding Sturm-Liouville problem. 

Figure \ref{Setup}(b) presents the fluid deformation given by Eq.(\ref{SolThinFilm}), showing that in the limit $w \ll d$, the length scale that governs $\eta$ is set by the width of the slot $d$. Consequently, we can approximate the change of the dielectric constant Eq.(\ref{epsilonD}) along the metal slab, by the value of the deformation at the center, $\eta(d/2,t \rightarrow \infty)$. In this limit the nonlinearity is reduced to a local Kerr-like cubic nonlinearity and the index change, $\Delta n_{D}$ can be represented as either $\Delta n_{D} =  \alpha_{TC} \Delta T$ or $\Delta n_{D} =  n_{2} \vert E_{0} \vert^{2}$, resembling index changes invoked due to traditional thermo-optical and electro-optical effects, where \cite{SuppInfo}
\begin{equation}
	n_{2}=\frac{4}{\pi^{6}} \frac{\sigma_{T}}{\sigma_{0}} \frac{d^{4}}{h_{0}} \frac{\alpha_{th}^{m}}{k_{th}^{m}}, \quad 
    \alpha_{TC}=\dfrac{3b}{2\pi^{3}} \frac{\sigma_{T}}{\sigma_{0}} \frac{d^{4}}{ w^{2} h_{0}^{2}}.
\end{equation}
It is instructive to compare index changes invoked by the thermocapillary effect, $\Delta n_{TC}$, to changes triggered by traditional thermo-optical effect, $\Delta n_{TO} = \alpha_{TO}\Delta T$ where $\alpha_{TO}$ is the thermo-optical coefficient. Assuming the values given below Eq.(\ref{hierarchy}), and $\lambda=800$ nm, $\alpha_{TO}=10^{-4}$ K$^{-1}$, $w=5$ \textmu m, $d=15$ \textmu m, and utilizing Eq.(\ref{epsilonD}), yields $\alpha_{TC}/\alpha_{TO} \simeq 10^{5}$.
The latter indicates that similar index changes, $\Delta n_{TO}$ and $\Delta n_{TC}$, require much smaller temperature increase for the case of the thermocapillary effect. In practice, this ratio is expected to be smaller due to thermal radiation losses and thermal advection.

Interestingly, thin film deformation driven by the thermocapillary effect introduces substantial changes of both real and imaginary parts of the dielectric function depending on the dielectric properties of the liquid.
Utilizing perturbation expansion of Eq.(\ref{MatchingCondCombined}) in the local Kerr-like nonlinearity limit, we can express SPP momentum with leading correction as $\beta=\beta_{0} + \Delta \beta$, where $\Delta \beta = \Delta \epsilon_{D} \beta_{0}^{ } q_{d}(q_{d}^{2}+\beta_{0}^{2})/(2 \epsilon_{D}^{2}(q_{d}^{\prime}+q_{d}^{ }))$ \cite{marini2009amplification}, $q_{d}^{2}=\beta_{0}^{2}(1-\epsilon_{D})$ and $q_{d}^{\prime}=\text{Re}(q_{d})$. The corresponding phase change $\Delta \beta$ vanishes for $\beta_{0}=0$ and tends to $- \Delta \epsilon_{D} \omega_{p}/(2 (1+\epsilon_{D})^{3/2})$ as $\beta_{0} \rightarrow \infty$, where $\omega_{p}$ is the plasma frequency in the metal. In particular, in case $\text{Im}(\epsilon_{D})>0$, 
changes of fluid thickness lead to power dependent changes of the real part of the depth averaged index as well as enhanced gain proportional to $\text{M}$.
This is inherently different than other gain mechanisms such as the electronic nonlinearity resulting from one- or two-photon processes \cite{boyd2003nonlinear}.

\begin{figure}[ht!]
	\includegraphics[scale=0.195]{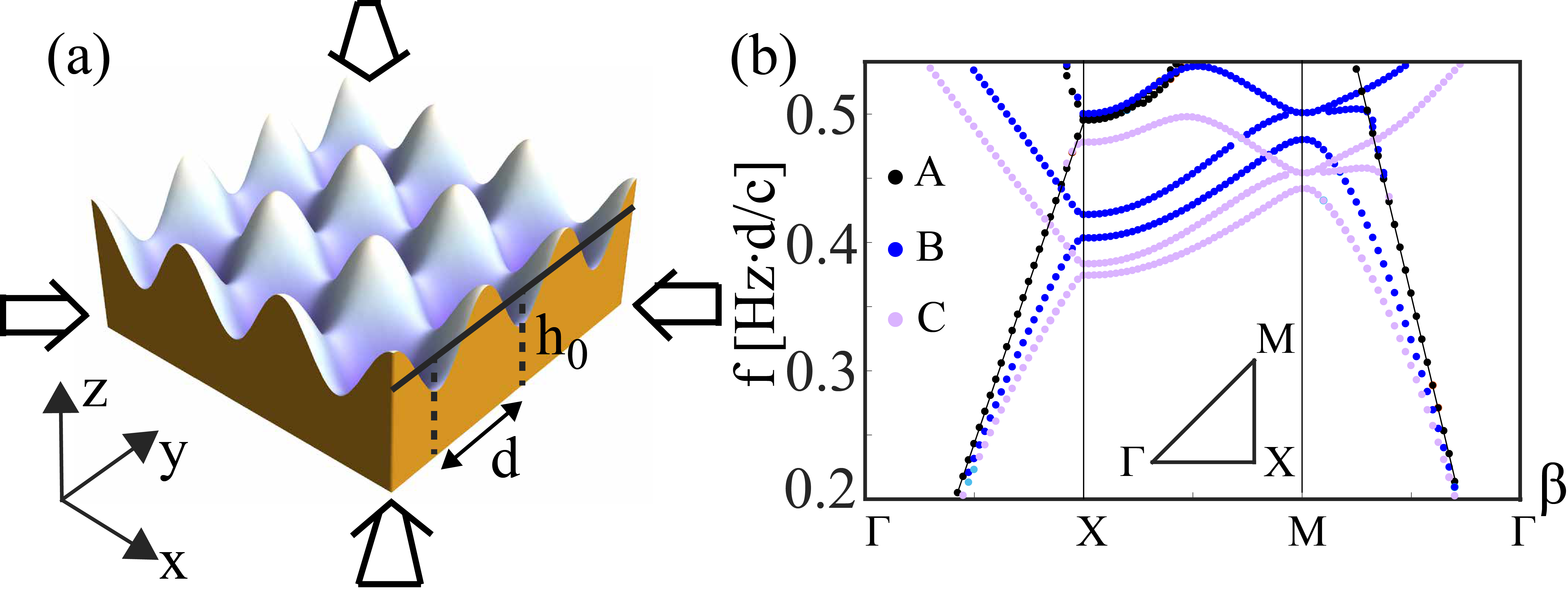}
    \caption{(a) Surface deformation given by Eq.(\ref{PeriodicDeformation2D}), optically induced by four SPPs (directions are indicated by arrows). (b) Band structure diagrams of the emerging FPPC for the cases: (B) $h_{0}=200$ nm and (C) $h_{0}=150$ nm. (A) presents the case without a liquid. Complete band gap occurs for $n=3$ (not shown), which is beyond the reach of current technology.}
    \label{Bandstructure}
\end{figure}

Consider now the case of self-induced change in the dispersion relation due to four SPP beams that propagate on a metal slab of width $w$, which has a size similar to the fluidic slot $d$ and satisfy $\lambda \ll w,d$. Furthermore, assume that these waves admit equal amplitude $\vert E_{0} \vert^{2}$, propagate along the directions $(\pm \hat{y} \pm \hat{z})/\sqrt{2}$, and admit a wave front larger than the size of the fluidic cell. Treating these beams as a plane waves yields the following optically induced intensity distribution, 
$16 \vert E_{0} \vert^{2} \cos^{2}(\beta_{0}x/\sqrt{2}) \cos^{2}(\beta_{0}y/\sqrt{2})$, and upon inserting it into Eq.(\ref{GreenGreen22}) yields the following deformation \cite{SuppInfo}
\begin{equation}
	\frac{\eta(\vec{r}_{\parallel},\infty)}{h_{0}} = - \dfrac{\tau_{l}}{16 \tau_{th} \lambda_{N}} \text{M} \cdot \cos \left( \beta_{0} x /\sqrt{2} \right) \cos \left( \beta_{0} y/\sqrt{2}  \right),
\label{PeriodicDeformation2D}    
\end{equation}
shown in Fig.\ref{Bandstructure}(a), where $N$ is an integer given by $N = \beta_{0} d/(\sqrt{2} \pi)$ and $\lambda_{n}$ is a constant given in Ref. \cite{SuppInfo}. The deformation described by Eq.(\ref{PeriodicDeformation2D}) admits a discrete translation symmetry along the $y,z$ axes, and constitutes an FPPC for a lower power SPP that propagates in this background over distances lower than its decay length. Specifically, Fig.\ref{Bandstructure}(b) presents the projection of the photonic band structure on the surface Brillouin zone for two different cases with different thicknesses $h_{0}=150$ and $h_{0}=200$ nm and periodicity $500$ nm, obtained by utilizing a commercial-grade simulator based on the finite-difference time-domain method \cite{Lumerical}.
To maximize the effect of the liquid, we have chosen the liquid index of value $n_{l}=2$, which is slightly below the index of selenium monobromide with $n_{l}=2.1$ \cite{meyrowitz1955compilation}.

\textbf{Nonlocal effects:}
Consider a single SPP beam of a finite spatial width, $\sigma_{SPP}$, that begins to propagate at $t=0$ along an infinite metal surface covered with a thin liquid film. The response of the thin film can be determined by the corresponding Green's function, $G_{l}$, of Eq.(\ref{ThinFilmEqU33}) with a source term $\delta(y)f(t)$. 
Film dynamics can be probed by considering an exponentially relaxing source with $f(t)=e^{-t/t_{s}}$, where $t_{s}$ is the relaxation time scale. 
The corresponding Green's function that vanishes in the limit of large times is given by \cite{SuppInfo}
\begin{equation}
	G_{l}(y)=-\frac{1}{D_{\sigma}^{ }} e^{-\vert y \vert/l_{s}} \Big[ \cos( \vert y \vert/l_{s} )+\sin( \vert y \vert/l_{s} ) \Big],
\label{GreensRelaxing}    
\end{equation}
where $l_{s}$ is the corresponding length scale given by $l_{s}=(4 D_{\sigma} t_{s})^{1/4}$ that can be tuned by choosing sufficiently large decay time scale $t_{s}$. Importantly, the Green's function Eq.(\ref{GreensRelaxing}) admits Taylor expansion at the origin and therefore allows to implement the Snyder-Mitchell model \cite{snyder1997accessible}, applicable for the strongly nonlocal regime. Expanding the Green's function inside the integral, Eq.(\ref{Schrodinger2}), as $G^{(0)}_{l}+\frac{1}{2} (y-y^{\prime})^{2}\partial^{2} G_{l}^{(0)}/\partial y^{2}$ yields a local  Schr{\"o}dinger equation with a harmonic oscillator potential
\begin{equation}
	2i\beta_{0} \dfrac{\partial \psi}{\partial z} = - \dfrac{\partial^{2} \psi}{\partial y^{2}} - \tilde{\chi}_{TC} I^{(0)} \dfrac{\partial^{2} G_{l}^{(0)}}{\partial y ^{2}} y^{2} \psi,
\label{SnyderMitchellLimit}    
\end{equation}
where $\psi=e^{-i z \tilde{\chi}_{TC} I^{(0)} \beta_{0}^{*} /(2 \vert \beta_{0} \vert^{2}) }A$, $\partial^{2}G_{l}^{(0)}/\partial y^{2}=2/(D_{\sigma}^{ }  l_{s}^{2})$, $I^{(0)}$ is the integral of $\vert A \vert^{2}$ along the $yz$ plane and $\beta_{0}^{*}$ is the complex conjugate of $\beta_{0}$. Notably, the sign of the potential term is determined by the sign of the Marangoni constant, and $\tilde{\chi}_{TC}<0$ simultaneously guarantees a Gaussian soliton \cite{SuppInfo}, analogous to the solution obtained in \cite{snyder1997accessible}, and an exponentially damping factor, $e^{z \tilde{\chi}_{TC} I^{(0)} \text{Im}(\beta_{0})/(2 \vert \beta_{0} \vert^{2}) }$, along the propagation direction which captures dissipation effects. 

\begin{figure}[ht!]
	\centering
	\includegraphics[scale=0.12]{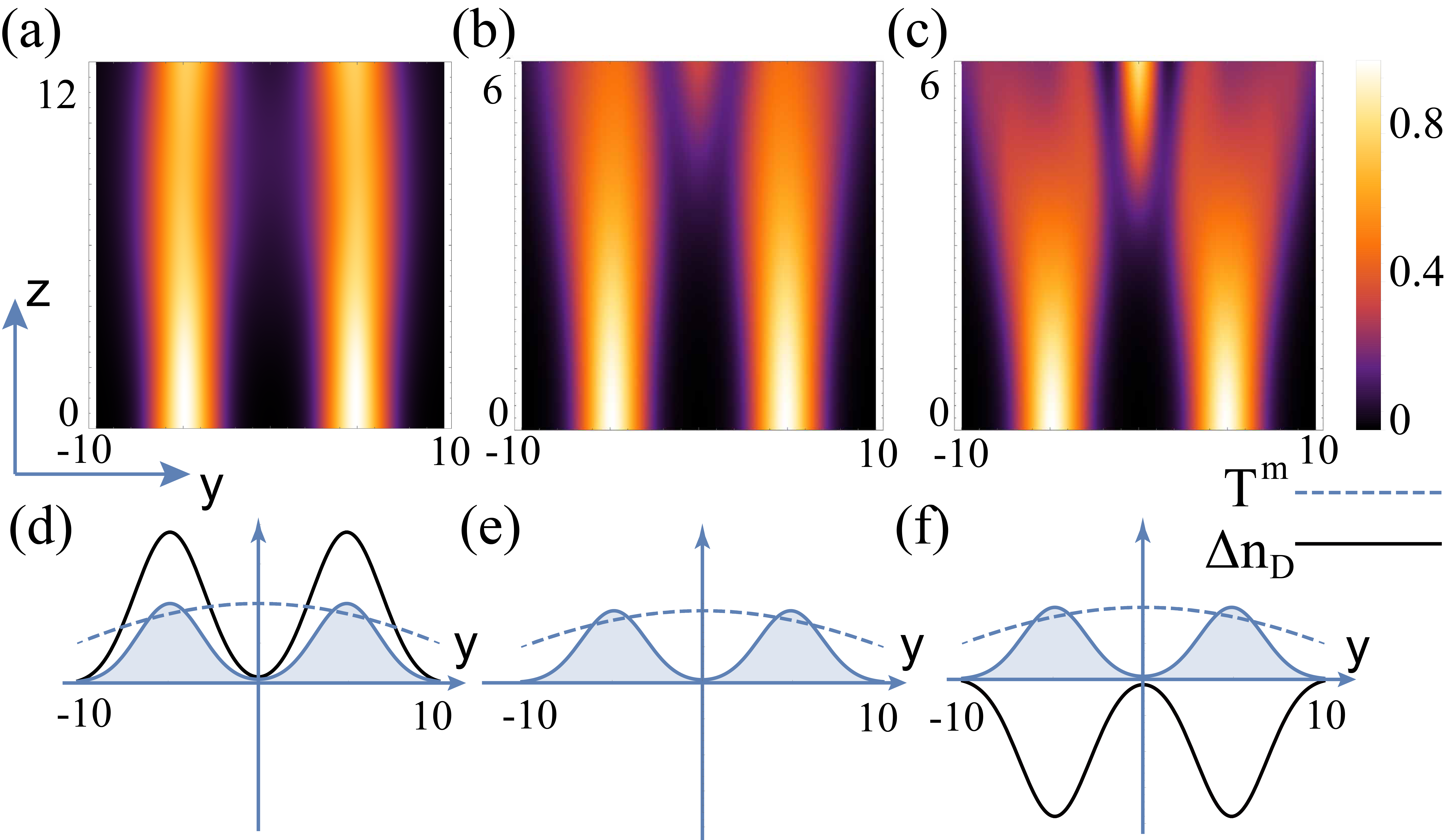}
    \caption{Numerical results presenting diffraction pattern of two SPPs for: (a) $\sigma_{T}<0$, (b) $\sigma_{T}=0$, and (c) $\sigma_{T}>0$. (d) ,(e) ,(f) present thin film shape (solid line), the optical intensity at $z=0$ (filled line) and the corresponding temperature distribution (dashed line) \cite{SuppInfo}. Relevant parameters: $\sigma_{SPP}=2.5/\sqrt{2}$, $\eta/h_{0}$=0.2, $\lambda=800$ nm, $h_{0}=200$ nm, $n_{l}=2$, $n_{g}=1$; coordinate axes normalized with respect to $k_{0}$.}
    \label{YoungInterference}
\end{figure}

To demonstrate the effect of nonlocality in case the correlation length due to the thermocapillary effect is comparable to $\sigma_{SPP}$, we turn to \textcolor{black}{commercial numerical solver \cite{Mathematica} and implement the built-in explicit Runge Kutta method}. Figure \ref{YoungInterference} presents an interference pattern of two parallel \textcolor{black}{SPP Gaussian beams of spatial variance $\sigma_{SPP}$}, analogous to Young's double-slit experiment in a leading order of a small parameter $\chi_{TC}$. In this approximation, the effect of the nonlocal self-induced spatial index change is taken into account by evaluating the integral in Eq.(\ref{Schrodinger2}) along the input beam [see Figs.(\ref{YoungInterference}d), (\ref{YoungInterference}e), (\ref{YoungInterference}f)] \cite{SuppInfo}. Figures \ref{YoungInterference}(a), \ref{YoungInterference}(b), \ref{YoungInterference}(c) present diffraction patterns of two SPPs due to the negative, zero, and positive Marangoni constant, respectively, that lead to induced dielectric function (empty graph) and temperature (dashed line). Notably, the index gradients required to support the self-focusing and defocusing effects in \ref{YoungInterference}(a) and \ref{YoungInterference}(c), respectively, are determined by the gradients of the optical beams intensities whereas the temperature field gradient is set by the much larger size of the metal slab. The self-focusing effect presented in Fig.(\ref{YoungInterference}a) yields solitary wave solutions where the deformed liquid acts as a waveguide for SPP beams. 

\textbf{Summary and concluding remarks:}
We presented a theoretical and numerical analysis of a novel thermocapillary self-induced, nonlocal and nonlinear mechanism for SPP-fluid interaction. In contrast to the traditional thermo-optical effect where the dielectric function modulation stems from changes of material density and polarization, the thermocapillary effect induces changes of the geometrical shape of the thin film. This generates waveguide-like structures in the fluid film, and exhibits much longer response time and correlation length than other nonlocal mechanisms reported to date.  
\textcolor{black}{The coupling described in this work is readily applicable} to other optical systems with heat dissipation such as photonic waveguides \cite{padmaraju2014resolving} and more general fluidic systems such as fluid-fluid interfaces, \textcolor{black}{and may be applicable for thermal imaging}. 
\textcolor{black}{Furthermore, dynamic modulation of other basic properties of the optical lattice such as symmetry and periodicity, opens a door to utilize our plasmonic system as a quantum simulator of a many body quantum systems such as topological insulators \cite{haldane1988model} recently realized in photonic systems \cite{rechtsman2013photonic, hafezi2013imaging}.}

\begin{acknowledgments}
\textbf{Acknowledgments:} S.R. cordially thanks Brandon Hong, Dr. Shiva Shahin and Dr. Valeri Frumkin for fruitful discussions.
This work was supported by the Office of Naval Research (ONR), Multidisciplinary University Research Initiative
(MURI), the National Science Foundation (NSF) (Grants No. DMR-1707641, No. CBET-1704085, No. ECCS-1405234, No. ECCS-1644647, No. CCF-1640227 and No. ECCS-1507146), the NSF ERC CIAN, the Semiconductor Recearch Corporation (SRC), the Defense Advanced Research Projects Agency (DARPA),  NSF's NNCI San Diego Nanotechnology Infrastructure (SDNI),  the Army Research Office (ARO), and the Cymer Corporation. 
\end{acknowledgments}


\widetext
\begin{center}
\newpage
\title{SI}
\textbf{SUPPLEMENTAL MATERIAL \\
Nonlocal and nonlinear surface plasmon polaritons and optical spatial solitons induced by the thermocapillary effect}

\text{Shimon Rubin and Yeshaiahu Fainman}

\textit{Department of Electrical and Computer Engineering, University of California, San Diego, 9500 Gilman Dr., La Jolla, California 92023, USA}
\end{center}

\setcounter{equation}{0}
\setcounter{figure}{0}
\setcounter{section}{0}
\setcounter{table}{0}
\setcounter{page}{1}

\renewcommand{\thesection}{S.\arabic{section}}
\renewcommand{\thesubsection}{\thesection.\arabic{subsection}}
\makeatletter 
\def\tagform@#1{\maketag@@@{(S\ignorespaces#1\unskip\@@italiccorr)}}
\makeatother
\makeatletter
\makeatletter \renewcommand{\fnum@figure}
{\figurename~S\thefigure}
\makeatother
\makeatletter \renewcommand{\fnum@table}
{\tablename~S\thetable}
\makeatother

\section{Derivation of the thin film Eq.(5)}

Consider the Navier-Stokes equations for a non-compressible fluid, given by
\begin{equation}
	\rho \left( \partial_{t}u_{i}+ u_{j} \partial_{j} u_{i} \right) = \partial_{j} \tau_{ij}+f_{i}; \quad i,j=x,y,z
\label{NavierStokes}    
\end{equation}
and subject to the incompressibility condition
\begin{equation}
	\partial_{i} u_{i}=0.
\label{DivU0}    
\end{equation}
Here, $f_{i}$ is the total force given as the sum $f_{i}=f^{b}_{i} + f^{s}_{i}$, where $f^{b}_{i}$ and $f^{s}_{i}$ are the body and the surface forces, respectively, whereas $\tau_{ij}=-p\delta_{ij}+\mu \left( \partial_{i}u_{j}+\partial_{j}u_{i} \right)$ is the stress tensor of the fluid with a dynamic viscosity $\mu$, mass density $\rho$ and pressure distribution $p$. 
We focus on shallow geometries and small Womersley (\text{Wo}) and reduced Reynolds numbers ($\epsilon$\text{Re}), constituted by 
the following conditions
\begin{equation}
\begin{split}
	\varepsilon=\dfrac{h_{0}}{d}  \ll 1; \quad \text{Wo}=&\dfrac{\tau_{diff}}{\tau_{f}}=\dfrac{\rho h_{0}^{2}}{\mu \tau_{f}} \ll 1; \quad \varepsilon \text{Re}=\dfrac{\tau_{diff}}{\tau_{adv}}=\dfrac{\rho u^{0}_{\parallel} h_{0}^{2}}{\mu r_{0}} \ll 1; 
\\
	&\tau_{diff}=\dfrac{\rho h_{0}^{2}}{\mu}; \quad \tau_{adv}=\dfrac{d}{u_{\parallel}^{0}},
\end{split}
\end{equation}
where $\text{Re}=\rho  u^{(0)} h_{0}/\mu_{0}$ is the Reynolds number, $\tau_{f}$ is the typical time-scale in the fluid due to external forcing, $\tau_{adv}$ is the advection time scale, and $h_{0}$, $u_{\parallel}^{0}$ and $d$ are the typical thin film depth, magnitude of in-plane fluid velocity and the scale of in-plane non-homogeneity (e.g. scale of surface tension variation), respectively.
We assume that the size of the system, $L$, is much larger than the typical width of the deformation, $\eta$, which together with the small deformation assumption combines to 
\begin{equation}
	 \eta \ll h_{0} \ll d \ll  L.
\end{equation}
In the leading order of $\epsilon$Re$\ll 1$ and Wo$\ll 1$, we can neglect the inertial terms, while in the limit of shallow geometry, $\varepsilon \ll 1$, which is equivalent to, 
$\partial / \partial r_{\parallel} \ll \partial/ \partial r_{\perp}$, we may drop the derivatives with respect to in-plane coordinates 
($\vec{r}_{\parallel}$) relative to the derivative with respect to the normal coordinate ($r_{\perp}$).  Under these assumptions, the leading order terms of the three dimensional Navier-Stokes equations reduce to the following 2D vector equation and a single 1D scalar equation, given by 
\begin{subequations}
\begin{align}
	\vec{\nabla}_{\parallel} p-\mu & \dfrac{\partial^{2} \vec{u}_{\parallel}}{\partial r^{2}_{\perp}}-\vec{f}_{\parallel}^{b} = 0
\\
	&\dfrac{\partial p_{ }}{\partial r_{\perp}}= f_{\perp}^{b},
\end{align}
\label{HSscaledLeading}
\end{subequations}
where $\vec{f}_{\parallel}^{b,s}$ and $\vec{f}_{\perp}^{b,s}$ stand for the body/surface forces along the tangent or the normal directions, respectively, given by $\vec{f}_{\parallel}^{b,s} =(\vec{f}^{b,s} \cdot \hat{t}) \hat{t} $ and $\vec{f}_{\perp}^{b,s} =(\vec{f}^{b,s} \cdot \hat{n}) \hat{n}$.
Next, we integrate equation of continuity, Eq.(\ref{DivU0}), use integration by parts, utilize the no penetration condition, $u_{\perp}=0$ on the bottom surface, $r_{\perp}=0$, and $u_{\perp}=\partial h/\partial t + \vec{u}_{\parallel} \cdot \nabla_{\parallel} h $ on the free interface, $h(\vec{r}_{\parallel},t) = h_{0}+\eta(\vec{r}_{\parallel},t)$, leading to
\begin{equation}
	\dfrac{\partial \eta}{\partial t} + \vec{\nabla}_{\parallel} \cdot \int\limits_{0}^{h} \vec{u}_{\parallel} dr_{\perp} = 0.
\label{IntegralContinuityEq}    
\end{equation}
The velocity field, $\vec{u}_{\parallel}$, is determined from integration of Eq.(\ref{HSscaledLeading}) supplemented with the following boundary conditions on the free surface
\begin{subequations}
\begin{align}
	n_{i} \tau_{ij} n_{j} = \sigma (\vec{\nabla} \cdot \hat{n}) + f_{\perp}^{s}
\\
	n_{i} \tau_{ij} t_{j} = \vec{\sigma} \cdot \hat{t} + f_{\parallel}^{s}.
\end{align}
\label{BCsurfaceTension}
\end{subequations}
Here, $\hat{t}$ and $\hat{n}$ are unit tangent and normal to the free interface, whereas $\sigma$ is the surface tension.
Following \cite{oron1997long2}, the leading order terms of Eq.(S\ref{BCsurfaceTension}) in $\epsilon$ and $C \epsilon^{-3}$ where $C$ is the capillary number, $C=u_{\parallel}^{0} \mu/\sigma_{0}$, are
\begin{subequations}
\begin{align}
	-p= \sigma \nabla^{2}_{\parallel} h + f_{\perp}^{s}
\\
	\mu \partial \vec{u}_{\parallel}/ \partial r_{\perp} = \vec{\nabla}_{\parallel} \sigma + \vec{f}_{\parallel}^{s}.
\end{align}
\label{BCsurfaceTensionLimit}
\end{subequations}
Integrating Eq.(S\ref{IntegralContinuityEq}), combining the result with Eq.(S\ref{BCsurfaceTensionLimit}) and assuming Navier boundary conditions with slip length, $1/\beta$, on the bottom surface, $r_{\perp}=0$, yields the following governing equation for the free surface $r_{\perp}=h(\vec{r}_{\parallel},t)$
in the presence of a body force potential $\varphi^{(b)}$ (e.g. gravitational potential) as well as surface and body non-potential forces
\begin{equation}
	 \dfrac{\partial h}{\partial t}+ \vec{\nabla}_{\parallel} \cdot \left( \dfrac{1}{\mu} \Bigg[  \left( \vec{f}_{\parallel}^{s} +\vec{\nabla}_{\parallel} \sigma \right) \left( \dfrac{1}{2} h^{2} + \beta h \right) - 
     \left( \dfrac{1}{3} h^{3} + \beta h^{2} \right) \left( \vec{\nabla}_{\parallel} \big[ \varphi^{(b)} - \sigma_{0} \nabla_{\parallel}^{2} h - f_{\perp}^{s} \big] + \vec{f}_{\parallel}^{b} \right) \Bigg] \right) =0,
\label{ThinFilmEqU}
\end{equation}
where $\sigma_{0}$ is the leading order term in power series in $\epsilon$.

Assume that the liquid rests on top of a metal film that is thin enough so one can neglect temperature variations along its vertical cross section and assume that the heat distribution in the metal is governed by a 2D heat equation
\begin{equation}
	\dfrac{\partial T^{m}}{\partial t} - D_{th}^{m} \nabla^{2}_{\parallel} T^{m} = \dfrac{\alpha^{m}_{th} D_{th}^{m}}{k_{th}^{m}} \vert E \vert^{2}.
\label{HeatMetal}    
\end{equation}
Here, $k_{th}^{m}$ and $D_{th}^{m}$ stand for the heat conductivity and heat diffusivity in the metal, respectively.
The temperature distribution in the liquid is governed by the following 3D diffusion equation (with $\nabla^{2}$ standing for a 3D Laplacian) 
\begin{equation}
	\dfrac{\partial T}{\partial t} - D_{th}^{ } \nabla^{2} T = 0,
\label{HeatEq3D}    
\end{equation}
where we assumed a spatially uniform heat conductance, as well as low thermal Peclet number, $\text{Pe}=\vert (\vec{u} \cdot \vec{\nabla}) T \vert/\vert D_{th} \nabla^{2} T \vert \sim u L /D_{th}$, which represents the ratio between advection of temperature due to liquid velocity $\vec{u}$ and thermal diffusion. The matching condition on the metal/liquid interface, $S$ dictates $T^{m} \vert_{S} = T \vert_{S}$.

In the shallow geometry approximation the in-plane derivatives are negligible compared to the derivatives along the normal direction, and 
the Eq.(S\ref{HeatEq3D}) takes the form
\begin{equation}
	\dfrac{\partial T}{\partial t} - D^{ }_{th} \dfrac{\partial^{2} T}{\partial r_{\perp}^{2}} = 0. 
\label{HeatEqTFilm}    
\end{equation}
On the fluid-gas interface, $r_{\perp}=h_{0}+\eta(\vec{r}_{\parallel})$, we assume that the temperature distribution is governed by the Newton's cooling law, $k_{th} \vec{\nabla} T \cdot \hat{n} + \alpha_{th} (T-T^{g}_{0}) = 0$, 
which in the limit of small deformation of the interface takes the form
\begin{equation}
\begin{split}
	\left( \dfrac{\partial T}{\partial r_{\perp}} + \dfrac{\alpha_{tr}}{k_{th}^{ }} (T-T^{g}_{0}) \right) \Big \vert_{r_{\perp}=h} =0.
\end{split}
\label{NewtonCooling}
\end{equation}
Here, $\alpha_{tr}$ is the heat transfer coefficient describing the rate of heat transfer from the liquid to the ambient gas phase and $T^{g}_{0}$ is a constant that stands for a gas temperature.  

Integrating Eq.(S\ref{HeatEqTFilm}), utilizing Eq.(S\ref{NewtonCooling}) and assuming sufficiently slow changes of the temperature field on the solid, $T \vert_{r_{\perp}=0}= T^{m}(\vec{r}_{\parallel})$, yields the following temperature field
\begin{equation}
	T(\vec{r}_{\parallel},r_{\perp})=T^{m}(\vec{r}_{\parallel})- \dfrac{\text{Bi}}{1+\text{Bi}}\left( T^{m}(\vec{r}_{\parallel})-T^{g}_{0} \right) \dfrac{r_{\perp}}{h(\vec{r}_{\parallel})},
\label{SolWithNewtonBC}    
\end{equation}
where $\text{Bi} = \alpha_{tr} h_{0}/k_{th}$ is the Biot number. Assuming that the surface tension is a function of temperature, 
$\sigma(T)=\sigma_{0} - \sigma_{T} (T-T_{0})$, where $\sigma_{0}$ and $\sigma_{T}$ are constants, and utilizing Eq.(S\ref{SolWithNewtonBC}) yields the corresponding surface tension gradient on the free surface
\begin{equation}
	\vec{\nabla}_{\parallel} \sigma = -\sigma_{T} \left( \vec{\nabla}_{\parallel} T + \frac{\partial T}{\partial r_{\perp}} \vec{\nabla} h \right) \Big \vert_{z=h} =   -\sigma_{T} \dfrac{1}{1+\text{Bi}} \vec{\nabla}_{\parallel} T^{m}.
\label{Tgrad}    
\end{equation}
In the limit of small Biot number,
\begin{equation}
	\text{Bi} \ll 1,
\end{equation}
(indeed for typical values $\alpha_{th} = 100 \text{ Wm}^{-2} \text{K}^{-1} , k_{th} = 0.2 \text{ W} \text{m}^{-1} \text{K}^{-1}, h_{0} = 0.5 \mu$m the Biot number is $\text{Bi} = 2.5 \cdot 10^{-4}$), shallow geometry ($\varepsilon \ll 1$), small changes of the surface tension ($\Delta \sigma/\sigma \simeq \varepsilon \eta/d$), and vanishing slip-length $\beta=0$, the thin film equation takes the form
\begin{equation}
	\dfrac{\partial \eta}{\partial t} - D_{l} \nabla^{2}_{\parallel} \left(\eta  - \ell^{2}_{c} \nabla^{2}_{\parallel} \eta  \right) = \dfrac{1}{2 \mu} \sigma_{T} h_{0}^{2} \nabla_{\parallel}^{2} T^{m},
\label{ThinFilmEqU3}
\end{equation}
where $\ell_{c}=\sqrt{\sigma_{0}/(\rho g)}$ is the so-called capillary length. \cite{de2013capillarity2}
In case we include also non-retarded van der Waals interaction, which is expected to dominate gravity on length scales below $100$ nm, Eq.(S\ref{ThinFilmEqU3}) takes the form
\begin{equation}
	\dfrac{\partial \eta}{\partial t} - D_{l} \nabla^{2}_{\parallel} \left((1-A_{H})\eta  - \ell^{2}_{c} \nabla^{2}_{\parallel} \eta  \right) =  \dfrac{\sigma_{T} h_{0}^{2}}{2 \mu}  \nabla_{\parallel}^{2} T^{m},
\label{ThinFilmEqU333}
\end{equation}
where $A_{H}=A_{H}^{0}/(2 \pi \rho g h_{0}^{4})$ is the normalized Hamaker constant $A^{(0)}_{H}$ \cite{israelachvili2011intermolecular2}.
Note that a thin film described by Eq.(S\ref{ThinFilmEqU333}) without a temperature forcing term, describes a stable configuration with respect to small deformations.
Indeed, substitution of $\eta= \eta_{0}(t)e^{i \vec{k} \cdot \vec{r}_{\parallel}}$ into Eq.(S\ref{ThinFilmEqU333}) (with vanishing right hand side) yields 
\begin{equation}
	\dfrac{d \eta_{0}}{d t} + D_{l} \left( (1-A_{H}) k^{2} + \ell^{2}_{c} k^{4}  \right) \eta_{0} = 0,
\label{ThinFilmStability}
\end{equation}
and exponentially decaying solution for repulsive molecular interaction ($A_{H}<0$). For sufficiently strong attractive interaction ($A_{H} > 1$) the critical wavenumber is $k_{c} = \sqrt{A_{H}-1}/l_{c}$ and the corresponding fastest diverging, spinodal wavelength, $\lambda_{s}$, \cite{cahn1965phase2}
is given by $\lambda_{s}/\sqrt{2} =   2\pi /k_{c} $.



\newpage

\section{Estimation of relative magnitude of electrostrictive and thermocapillary effects}

The force density acting on a dielectric fluid in presence of an electromagnetic field is given by \cite{landau2013electrodynamics2}
\begin{equation}
	\vec{f}^{H}= \rho_{f} \vec{E} -\dfrac{1}{2}E^{2} \vec{\nabla} \epsilon + \dfrac{1}{2} \vec{\nabla} \left( \rho \dfrac{\partial \epsilon}{\partial \rho} E^{2} \right)+
   (\epsilon-1) \dfrac{\partial}{\partial t} (\vec{E} \times \vec{H}).
\label{HForce}
\end{equation}
Here, $\rho_{f}$ stands for possible surface charges on gas-fluid interface, the second term stems from changes of the dielectric constant, the third term corresponds to electrostrictive due to changes of density and the last term is the so-called Abraham term. Given the fact that the mechanical response time of the thin film is much smaller then period time of an oscillating optical wave, the first and the last term average to zero. Employing the Claussius-Mosotti relation for non-polar dielectrics (e.g. oil), $\rho \partial \epsilon/\partial \rho =\epsilon_{0} (\epsilon_{r}-1)(\epsilon_{r}+2)/3$, \cite{stratton2007electromagnetic2}  (or its analogue $\rho \partial \epsilon/\partial \rho = a \epsilon$ for polar dielectrics such as water, where $a<1.5$ holds for most of the studied polar dielectric liquids \cite{jacobs1952analysis2}), where $\epsilon_{r}=\epsilon/\epsilon_{0}$.
Assuming that the dielectric function doesn't change along the $\hat{t}$ direction yields
\begin{subequations}
\begin{align}
	\vec{f}_{\perp}^{s} & =- \dfrac{ \epsilon_{0} }{2}(\epsilon_{r} - 1)E^{2} 
    \\
 \vec{f}_{\parallel}^{b} &=  \dfrac{1}{2} \vec{\nabla}_{\parallel} \left(\rho \dfrac{\partial \epsilon}{\partial \rho} E^{2} \right) = \dfrac{\epsilon_{0}}{6} \vec{\nabla}_{\parallel} \left( (\epsilon_{r}-1)(\epsilon_{r}+2) E^{2} \right)
\label{SurfaceH},
\end{align}
\end{subequations}
which implies that these admit identical order of magnitude in Eq.(S\ref{ThinFilmEqU}).
The ratio of thermocapillary and dielectric terms scales as a large number
\begin{equation}
	\dfrac{\vert \frac{h_{0}^{2}}{2 \mu} \nabla^{2}_{\parallel} \sigma \vert}{\vert  \frac{h_{0}^{3}}{6 \mu} \epsilon_{0} (\epsilon_{r}-1) \nabla^{2}_{\parallel}I  \vert} = \dfrac{\vert \frac{h_{0}^{2}}{2 \mu} \frac{\sigma_{T} \alpha_{th} I}{k_{th}} \vert}{\vert \frac{h_{0}^{3}}{6 \mu} \epsilon_{0} (\epsilon_{r}-1) \frac{I}{d^{2}} \vert} \simeq \dfrac{3d}{h_{0}} \dfrac{r_{0} \sigma_{T} \alpha_{th}}{\epsilon_{0} k_{th}} \simeq  10^{5},
\end{equation}
provided both originate from a source with identical intensity, $I$, and where in the last step we utilized the following values: $d=200$ nm, $h_{0}=400$ nm, $\sigma_{T}=10^{-4}$ Nm$^{-1}$K$^{-1}$, $\epsilon_{0}=8.85 \cdot 10^{-12}$ Fm$^{-1}$, $k_{th}=3 \cdot 10^{2}$ Wm$^{-1}$K$^{-1}$ and $\alpha_{th}=7.7 \cdot 10^{7}$ m$^{-1}$.

\newpage

\section{Derivation of Eq.(5)}

The coupled processes of the heat transport and the film deformation are described, respectively, by Eq.(S\ref{HeatMetal}) and Eq.(S\ref{ThinFilmEqU3}), whereas the corresponding Green's functions are defined by the following relations
\begin{equation}
\begin{split}
	\dfrac{\partial G_{th}}{\partial t} - D_{th}^{m} \nabla^{2}_{\parallel} G_{th} = \delta(\vec{r}_{\parallel})\delta(t), 
\\
	\dfrac{\partial G_{l}}{\partial t} - D_{l} \nabla^{2}_{\parallel} \left(G_{l} - \ell^{2}_{c} \nabla^{2}_{\parallel} G_{l}  \right) = \delta(\vec{r}_{\parallel})\delta(t).
\end{split}
\end{equation}
Formally, the deformation admits the following representation in terms of the convolution of the two Green's functions
\begin{equation}
	\dfrac{\eta (\vec{r}_{\parallel},t)}{h_{0}} = \dfrac{1}{2} \text{Ma} \cdot \chi \cdot \dfrac{1}{\tau_{th}^{2}} 
    \int d\vec{r}^{\prime}_{\parallel} d\vec{r}^{\prime \prime}_{\parallel} dt^{\prime} dt^{\prime \prime} G_{l}(\vec{r}_{\parallel}-\vec{r}^{\prime}_{\parallel},t - t^{\prime}) 
    \nabla^{\prime 2}_{\parallel} G_{th}(\vec{r}^{\prime}_{\parallel}-\vec{r}^{\prime \prime}_{\parallel},t^{\prime} - t^{\prime \prime}) \frac{I(\vec{r}_{\parallel}^{\prime \prime},t^{\prime \prime})}{I_{0}} \Big \vert_{\vec{r}^{\prime}_{\parallel}=(\vec{r}_{\parallel},h(\vec{r}_{\parallel}))},
\label{GreenGreen}    
\end{equation}
where, $I=\vert E(\vec{r}^{\prime}_{\parallel}) \vert ^{2}$ is the field intensity, $\nabla^{\prime 2}_{\parallel}=(\partial^{2}/\partial(x/d)^{2},\partial^{2}/\partial(y/r_{0})^{2})$ denotes the Laplacian operator with respect to the normalized in-plane primed coordinates, and $\tau_{th} = d^{2}/D_{th}^{m}$. Note that for sufficiently long times, $\tau_{th} \ll t \ll \tau_{l}$, we can consider the thermal distribution as quasi static and consequently subject to Eq.(S\ref{HeatMetal}) without the time derivative term. 
In such case, $-D_{th}^{m} G_{th}$ is the Green's function of the Laplacian operator and the expression given by Eq.(S\ref{GreenGreen}) simplifies to
\begin{equation}
	\dfrac{\eta (\vec{r}_{\parallel},t)}{h_{0}} = - \dfrac{1}{2} \text{Ma} \cdot \chi \cdot \dfrac{1}{\tau_{th}} 
    \int d\vec{r}^{\prime}_{\parallel} dt^{\prime}  G_{l}(\vec{r}_{\parallel}-\vec{r}^{\prime}_{\parallel},t - t^{\prime}) 
    \frac{I(\vec{r}^{\prime}_{\parallel},t^{\prime})}{I_{0}} \Big \vert_{\vec{r}^{\prime}_{\parallel}=(\vec{r}_{\parallel},h(\vec{r}_{\parallel}))}.
\end{equation}

\section{Green's function for thin film deformation in a rectangular slot}

Consider a 1D case where the thin film Eq.(S\ref{ThinFilmEqU33}) takes the following form 
\begin{equation}
	\dfrac{\partial \eta_{G}}{\partial t} - A \dfrac{\partial^{2} \eta_{G}}{\partial x^{2}} + B \dfrac{\partial^{4} \eta_{G}}{\partial x^{4}} = \delta(x-x_{0}) \delta(t); \quad A= - \dfrac{\rho g h_{0}^{3}}{3 \mu}, B= - \ell_{c}^{2} A.
\label{GreenThinFilm}    
\end{equation}
Assume that the liquid is bound by vertical walls at $x=0,d$ and is subject to vanishing fluxes on the walls, constituted by $\partial \eta/\partial x \vert_{x=0,d}=0$ and $\partial^{3} \eta/\partial x^{3} \vert_{x=0,d}=0$.
The latter physically corresponds to liquid which forms a $\pi/2$ wetting angle with the walls at $x=0$ and $x=d$, and satisfies volume conservation condition, i.e. vanishing of the integral of $\eta_{G}$ from $x=0$ to $x=d$.
The corresponding solution, $\eta_{G}$, of this Sturm-Liouville problem can be found by employing the closure relation $ \sum\limits_{j=1}^{\infty} \varphi_{j}(x) \varphi_{j}(x_{0}) = \delta(x-x_{0}) $ where $\varphi_{j}(x)$ is the corresponding $j$-th eigenfunction. Multiplying Eq.(S\ref{GreenThinFilm}) by $\varphi_{k}(x)=\sqrt{2/d}\cos(k \pi x/d)$ and integrating from $0$ to $d$ yields
\begin{equation}
	\eta_{G}(x,t)=\sum\limits_{j=1}^{\infty} \varphi_{j}(x_{0}) \varphi_{j}(x) e^{-\lambda_{n} t}; \quad 
    \lambda_{n} = \pi^{4} \Big[ n^{4} - \left( \dfrac{d}{ \pi \ell_{c}} \right)^{2} n^{2} \Big] \dfrac{1}{\tau_{l}}.
\label{Green1D}    
\end{equation}
Integrating Eq.(S\ref{Green1D}) with respect to time, yields
\begin{equation}
	\eta_{H}(x,t)=\sum\limits_{n=1}^{\infty} \dfrac{1}{\lambda_{n}} \varphi_{n}(x)  \varphi_{n}(x_{0}) \left(1 - e^{-\lambda_{n} t} \right),
\label{GreenEtaH}    
\end{equation}
which solves the equation Eq.(S\ref{GreenThinFilm}) with a non-homogeneous term $H(t)\delta(y-y_{0})$ (where $H(t)$ is the Heaviside step function).

Similarly the solution to a 2D problem 
\begin{equation}
	\dfrac{\partial \eta_{H}}{\partial t} - A \nabla^{2} \eta_{H} + B \nabla^{4} \eta_{H} = \delta(x-x_{0}) \delta(y-y_{0}) H(t),
\end{equation}
subject to boundary conditions $\vec{\nabla} \eta_{H} \cdot \hat{n} = 0$ at $x=0,d_{x}$ and $y=0,d_{y}$ as well as initial conditions $\eta_{H}(t=0,\vec{r}) =0$, is explicitly given by
\begin{equation}
	\eta_{H}(x,y,t) =  \sum\limits_{n,m=1}^{\infty} \cos \left( \dfrac{m \pi x_{0}}{d_{x}} \right) \cos \left( \dfrac{n \pi y_{0}}{d_{y}} \right) \dfrac{1}{\lambda_{m,n}} ( 1- e^{- \lambda_{m,n} t} ) \varphi_{m,n}(x,y).
\end{equation}
where
\begin{equation}
	\lambda_{m,n} = \pi^{2} \left( A \left( \dfrac{m^{2}}{d_{x}^{2}} + \dfrac{n^{2}}{d_{y}^{2}} \right) + B \pi^{2} \left( \dfrac{m^{4}}{d_{x}^{4}} + 6 \dfrac{m^{2}}{d_{x}^{2}} \dfrac{n^{2}}{d_{y}^{2}}  +\dfrac{n^{4}}{d_{y}^{4}} \right) \right).
\end{equation}
and $\varphi_{m,n}(x,y)$ are the corresponding eigenfunctions
\begin{equation}
	\varphi_{m,n}(x,y) = \dfrac{2}{d_{x} d_{y}} \cos \left( \dfrac{m \pi x}{d_{x}} \right) \cos \left( \dfrac{n \pi y}{d_{y}} \right).
\end{equation}

Applying a similar method for the heat transport equation, obtained from Eq.(S\ref{GreenThinFilm}) by setting $B=0$ and $A=D_{th}^{m}$, the temperature field $T_{H}$ that vanishes at $x=(d \pm w)/2$ and is subject to a vanishing initial temperature at $t=0$, is given by
\begin{equation}
	T_{H}(x,t) = \dfrac{w^{2}}{\pi^{2} D_{th}^{m}} \sum\limits_{n=1}^{\infty} \dfrac{1}{n^{2}} f_{n}(x) f_{n}(x_{0}) \left( 1 - e^{-\frac{n^{2} \pi^{2}}{\tau_{th}} t} \right).
\label{GreenTemperatureHeaviside}    
\end{equation}
Here, $\tau_{th}=w^{2}/D_{th}^{m}$ and the corresponding complete set of eigenfunctions is given by $f_{n}(x)=\sqrt{2/w}\cos \left( 2 \pi (n-1/2) (x-d/2)/w \right)$. The temperature distribution, $T^{m}(x,t)$, due to arbitrary source is given by convolving Green's function, Eq.(S\ref{GreenTemperatureHeaviside}), with a source $(D_{th}^{m}/k_{th}^{m})\alpha_{th}^{m} \vert E \vert^{2}$. For the case $E=E_{0} \cos(\pi (x-d/2)/w)$ where $x$ lays in the interval, $(d-w)/2<x<(d+w)/2$, the corresponding expression for the temperature of the metal (and the dielectric) is given by
\begin{equation}
	T^{m}(x,t)= \dfrac{8w^{2}}{3\pi^{3}} \dfrac{\alpha_{th}^{m} \vert E_{0} \vert^{2}}{k_{th}^{m}}  \cos \left( \dfrac{\pi (x-d/2)}{w} \right) \left( 1 - e^{- D_{th}^{m} \left( \dfrac{\pi}{w} \right)^{2} t} \right).
\label{Tm}    
\end{equation}
Other choices of the electric field intensity distrubution along the slab, only lead to a different numeric factor which multiplies the $w^{2}$ term in Eq.(S\ref{Tm}). The corresponding deformation is found by taking the following integral $\frac{1}{2 \mu}\sigma_{T}h_{0}^{2}\int\limits_{0}^{d} dx_{0} \eta_{H}(x-x_{0},t) \frac{\partial^{2}}{\partial x_{0}^{2}} T^{m}(x_{0},\infty)$, leading to
\begin{equation}
	\dfrac{\eta(x,t)}{h_{0}} =\dfrac{16}{\pi^{6}} \dfrac{\sigma_{T}}{\sigma_{0}} \dfrac{w d^{3}}{ h_{0}^{2}} \dfrac{\alpha_{th}^{m} \vert E_{0} \vert^{2}}{k_{th}^{m}} \sum\limits_{n=1}^{\infty} (-1)^{n}
    \dfrac{\cos \left( \dfrac{n \pi w}{2d} \right)}{1-\left( \dfrac{n w}{d} \right)^{2}}    
    \dfrac{\cos \left( \dfrac{n \pi x}{d} \right)}{n^{4} - \left( \dfrac{d}{ \pi \ell_{c}} \right)^{2} n^{2} } \left( 1 - e^{-\pi^{4} \Big[ n^{4} - \left( \dfrac{d}{ \pi \ell_{c}} \right)^{2} n^{2} \Big] \dfrac{t}{\tau_{l}}} \right).
\label{SolThinFilm5}    
\end{equation}
Note, that this expression is valid at times $t>\tau_{th}$ (so one can omit the last term in Eq.(S\ref{Tm})) and that the viscosity appears only in the time scale, $\tau_{l}$, as could be also anticipated by considering the static limit of Eq.(S\ref{ThinFilmEqU33}).

\section{Green's function for thin film deformation without confining walls}

\subsection{Exponentially decaying solution in time} 

\begin{figure}[th]
	\centering
\includegraphics[scale=0.35]
  {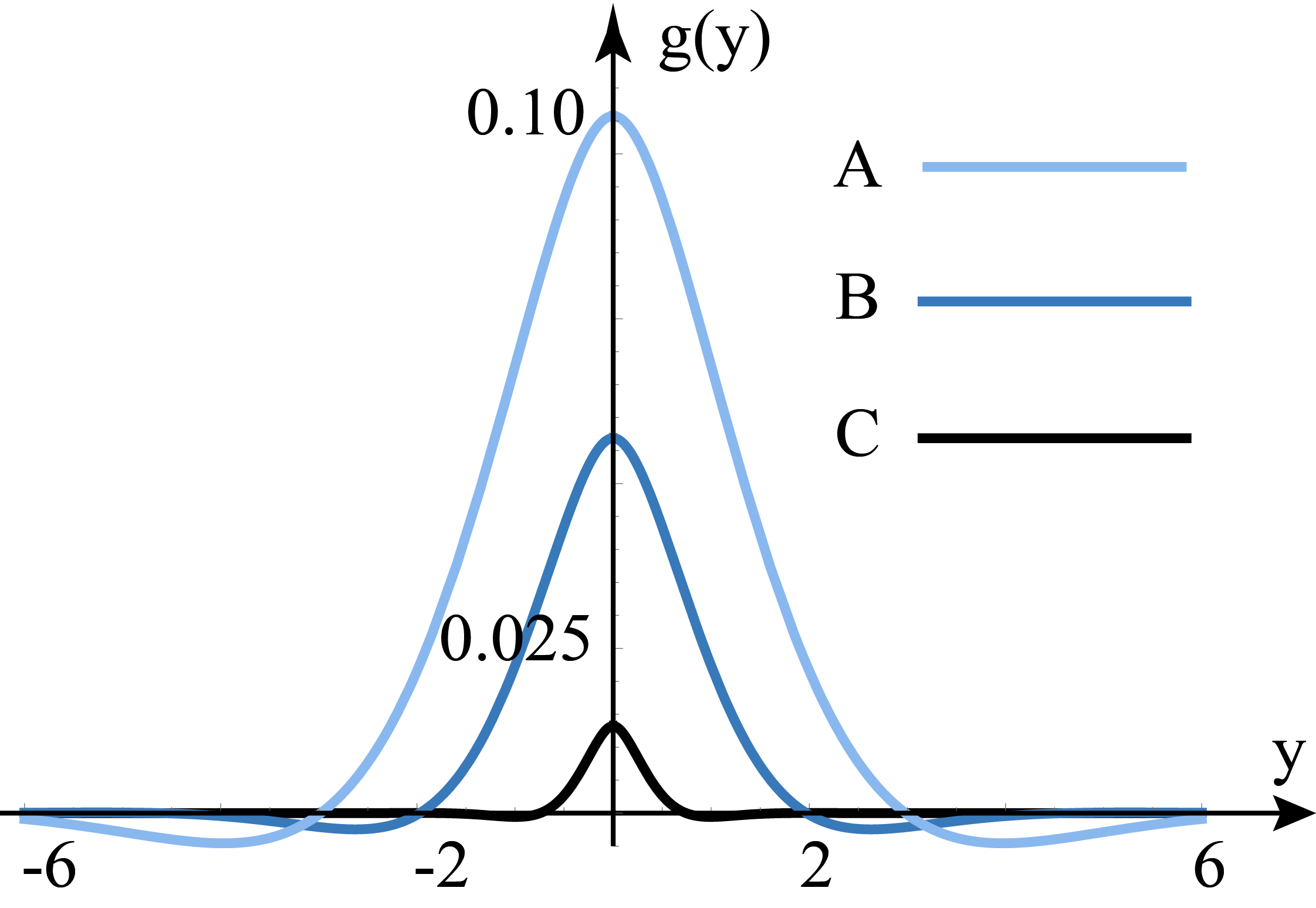}
    \caption{Green's function solution given by Eq.(S\ref{GreenExp}) for three different values of the relaxation time scale $t_{s}$: (A) $t_{s}=0.8$, (B) $t_{s}=0.35$ and (C) $t_{s}=0.05$. Larger time scales correspond to wider spatial profiles of the deformation.}
    \label{GreenExp}
\end{figure}

Thin film equation in 1D in dimensionless variables is given by
  \begin{equation}
	\dfrac{\partial \eta}{\partial t} + \dfrac{\partial^{4} \eta} {\partial y^{4}} = \dfrac{3}{2} \dfrac{\sigma_{T} d^{4}}{\sigma_{0} h_{0}^{2}}   \dfrac{\alpha I}{k_{th}}.
\label{GreenExpDecayThinFilm}    
\end{equation}
Here, we have rescaled to dimensionless variables via $y \rightarrow d \cdot y$, $t \rightarrow d^{4}/(D_{\sigma}) t$, $\eta \rightarrow h_{0} \eta$, where $D_{\sigma} =\sigma_{0} h_{0}^{3}/(3 \mu)$.
The corresponding Green's function due to an exponentially decaying time dependent source is given by
\begin{equation}
	\dfrac{\partial \eta}{\partial t} + \dfrac{\partial^{4} \eta} {\partial y^{4}} = e^{-t/t_{s}} \delta(y),
\label{GreenExpDecayThinFilm}    
\end{equation}
where $t_{s}>0$. Assuming that $\eta$ admits a separable form $\eta(x,t)=e^{-t/t_{s} }g(y)$, the corresponding solution to the governing equation for $g(y)$
\begin{equation}
	-\dfrac{g}{t_{s}} + \dfrac{d^{4}g}{d y^{4}} = \delta(y),
\end{equation}
is given by (see Fig.(S\ref{GreenExp}))
\begin{equation}
	g(y)=\dfrac{1}{2 \pi} \int\limits_{0}^{\infty} \dfrac{\cos(ky)dk}{k^{4}-1/t_{s}} = -\dfrac{t_{s}^{3/4}}{8} e^{-y / \sqrt{2 t_{s}^{1/2}}} \left( \cos \left( y/ \sqrt{2 t_{s}^{1/2}} \right) + \sin \left( y/ \sqrt{2 t_{s}^{1/2}} \right)  \right) \quad \text{for } y \geq 0.
\label{gy}    
\end{equation}
The corresponding second derivative with respect to $y$ is given by
\begin{equation}
	\dfrac{d^{2}g(y)}{dy^{2}} =  \dfrac{t_{s}}{8} \left( \cos \left( y/ \sqrt{2 t_{s}^{1/2}} \right) - \sin \left(y/ \sqrt{2 t_{s}^{1/2}} \right)  \right).
\end{equation}

\subsection{Green's function of the time dependent transient problem} 

The Green's function, $\eta_{G}$, which satisfies the following nondimensional thin film equation
\begin{equation}
	\dfrac{\partial \eta_{G}}{\partial t} + \dfrac{\partial^{4} \eta_{G}} {\partial y^{4}} = \delta(t) \delta(y),
\end{equation}
can be represented as
\begin{equation}
	\eta_{G}(\vec{r}_{\parallel},t) = \dfrac{1}{2 \pi} \int\limits_{- \infty}^{\infty} e^{-k^{4}t}e^{ikx}dk,
\label{GreenIntegral1D}    
\end{equation}
where the summation along the $k-$axis reflects that there is no restriction on the choice of wave-vectors. Integration of Eq.(S\ref{GreenIntegral1D}) yields,
\begin{equation}
	\eta_{G}(y,t) = \dfrac{2 \Gamma(\frac{5}{4}) _{0}F_{2}(;\frac{1}{2},\frac{3}{4};\frac{x^{4}}{256t})}{t^{1/4}} - \dfrac{x^{2} \Gamma(\frac{3}{4}) _{0}F_{2}(;\frac{5}{4},\frac{3}{2};\frac{x^{4}}{256t})}{4 t^{3/4}},
\label{GreenActuation}    
\end{equation}
where $_{0}F_{2}$ is a Hypergeometric function $_{p}F_{q}$ with $p=0, q=2$.

\section{Derivation of Eq.(13)}

It is instructive to compare the contributions of thermocapillary and thermo-optical effects, $\Delta n_{TC}$ and $\Delta n_{TO}$, respectively, to the changes of the depth averaged index. The depth averaged index weighted by an exponential factor $2 q_{D} e^{-2 q_{D} x}$ is given by
\begin{equation}
	n_{D}(\eta(\vec{r}_{\parallel})) = 2q_{d} \int\limits_{0}^{\infty} n(x) e^{-2q_{d}x}dx = n_{l} +  (n_{g}-n_{l})e^{-2 q_{d} (h_{0} + \eta(\vec{r}_{\parallel}))}.
\label{IndexDepthAverage}    
\end{equation}
Here, $n_{l}$ and $n_{g}$ are the indices in the regions $0<x<h$ and $x>h$, respectively. Assuming the deformation is small relative to the decay length of the SPP into the bulk, and keeping the linear term in the Taylor expansion of the index  Eq.(S\ref{IndexDepthAverage}) with respect to $q_{d} \eta$ yields
\begin{equation}
n_{D}(\eta(\vec{r}_{\parallel})) = n_{0} + \Delta n_{D},
\end{equation}
where
\begin{equation}
	n_{0}= n_{l}+ (n_{g}-n_{l}) e^{-2q_{d}h_{0}} ; \quad \Delta n_{D} = b \eta(\vec{r}_{\parallel})/h_{0}; \quad b=\Big[ 2 q_{d} h_{0} (n_{l} - n_{g} ) \Big] e^{-2q_{d}h_{0}}.
\label{DepthAveragedIndex}    
\end{equation}
The corresponding dielectric function can be represented as
\begin{equation}
	\varepsilon_{D}(\eta(\vec{r}_{\parallel})) = \epsilon_{0} + \Delta \epsilon_{D}
\end{equation}
where
\begin{equation}
	\epsilon_{0} = n_{0}^{2}; \quad \Delta \epsilon_{D} = 2 n_{0} \Delta n_{D} .
\end{equation}
For the model where the temperature and the deformation are described by Eq.(S\ref{Tm}) and Eq.(S\ref{SolThinFilm5}), respectively, the index change, $\Delta n_{D}$ can be represented as either $\Delta n_{D} =  \alpha_{TC} \Delta T$ or $\Delta n_{D} =  n_{2} \vert E_{0} \vert^{2}$, resembling index changes invoked due to traditional thermo-optical and electro-optical effects. Utilizing Eq.(S\ref{DepthAveragedIndex}) and taking $\Delta T$ as the value of $T^{m}(x,t)$, given by Eq.(S\ref{Tm}), at $x=d/2$ and $t \rightarrow \infty$, yields
\begin{equation}
	\alpha_{TC}=\dfrac{3b}{2\pi^{3}} \frac{\sigma_{T}}{\sigma_{0}} \frac{d^{4}}{ w^{2} h_{0}^{2}}; \quad
	n_{2}=\frac{4}{\pi^{6}} \frac{\sigma_{T}}{\sigma_{0}} \frac{d^{4}}{h_{0}} \frac{\alpha_{th}^{m}}{k_{th}^{m}}. 
\end{equation}

\section{Derivation of eq.(10)}

The governing equations for SPP are Maxwell equations written below in dimensionless cartesisan coordinates $x,y,z$, scaled by $1/k_{0}=\lambda_{0}/(2 \pi)$, where $\lambda_{0}$ is the vacuum wavelength
\begin{subequations}
\begin{align}
	& \frac{\partial^{2} E_{y,\alpha}}{\partial x \partial y} - \frac{\partial^{2} E_{x,\alpha}}{\partial y \partial y} - \frac{\partial^{2} E_{x,\alpha}}{\partial z \partial z} + \frac{\partial^{2} E_{z,\alpha}}{\partial z \partial x} = \epsilon_{\alpha} E_{x,\alpha}
\\
	& \frac{\partial^{2} E_{z,\alpha}}{\partial y \partial z} - \frac{\partial^{2} E_{y,\alpha}}{\partial z \partial z} - \frac{\partial^{2} E_{y,\alpha}}{\partial x \partial x}  + \frac{\partial^{2} E_{x,\alpha}}{\partial x \partial y} = \epsilon_{\alpha} E_{y,\alpha}
\\
	& \frac{\partial^{2} E_{x,\alpha}}{\partial z \partial x} - \frac{\partial^{2} E_{z,\alpha}}{\partial x \partial x} - \frac{\partial^{2} E_{z,\alpha}}{\partial y \partial y}  + \frac{\partial^{2} E_{x,\alpha}}{\partial x \partial y} = \epsilon_{\alpha} E_{z,\alpha},
\end{align}
\label{Maxwell3D}
\end{subequations}
and the continuity of the normal component of the displacement and tangential components of the electric field
\begin{equation}
\begin{split}
	\Big[ \epsilon_{0d} + \text{M} \int d \vec{r}^{\prime}_{\parallel} G_{l}(\vec{r}_{\parallel},\vec{r}_{\parallel}^{\prime}) \left(  \vert E_{x}  \vert^{2} + \vert E_{y} \vert^{2} + \vert E_{z} \vert^{2} \right) \vert_{\vec{r}_{\parallel} = \vec{r}_{\parallel}^{\prime}} \Big] E_{x,\alpha} \vert_{x=0} = \epsilon_{m} E_{x,m} \vert_{x=0}
    \\
    E_{z,m} \vert_{x=0} = E_{z,d} \vert_{x=0}, \quad
    E_{y,m} \vert_{x=0} = E_{y,d} \vert_{x=0}.
\end{split}	
\end{equation}
Here, the index $\alpha$ runs over the values $\alpha=m,d$, that correspond to metal (m) and dielectric (d), which occupy the regions $x<0$ and $x>0$, respectively. The dielectric function in each region is given by
\begin{equation}
\begin{split}
	\epsilon_{m} &= \epsilon_{m}^{\prime} + i \epsilon_{m}^{\prime \prime}
\\
	\epsilon_{d} &= \epsilon_{0d} + \text{M} \int d \vec{r}^{\prime}_{\parallel} G_{l}(\vec{r}_{\parallel},\vec{r}_{\parallel}^{\prime}) \left(  \vert E_{x}  \vert^{2} + \vert E_{y} \vert^{2} + \vert E_{z} \vert^{2} \right) \vert_{\vec{r}_{\parallel} = \vec{r}_{\parallel}^{\prime}},
\end{split}
\end{equation}
where $\epsilon_{0d}$ can have both real and imaginary parts, $\epsilon_{0d} = \epsilon_{0d}^{\prime} + i \epsilon_{0d}^{\prime \prime} $.
Assume that the components of the electrical field admit the following expansion in powers of $\chi_{TC}$
\begin{equation}
\begin{split}
	E_{x,\alpha}^{} &= \Big[ A_{x,\alpha}^{(0)} + A_{x,\alpha}^{(1)} + o(\vert \text{M} \vert^{2})  \Big] e^{i \beta_{0} z}
\\
	E_{y,\alpha}^{} &= \Big[ A_{y,\alpha}^{(0)} + o(\vert \text{M} \vert^{3/2}) \Big] e^{i \beta_{0} z}
\\
	E_{z,\alpha}^{} &= \Big[ A_{z,\alpha}^{(0)} + A_{z,\alpha}^{(1)} + o(\vert \text{M} \vert^{2}) \Big] e^{i \beta_{0} z},
\end{split}
\end{equation}
and the following functions and their derivatives, respectively, scale as
\begin{equation}
\begin{split}
	A_{x,j}^{(0)}, A_{z,j}^{(0)} & \sim \vert \text{M} \vert^{0}
\\
	A_{y,j}^{(0)} & \sim \vert \text{M} \vert^{1/2}
\\
	A_{x,j}^{(1)}, A_{z,j}^{(1)} & \sim \vert \text{M} \vert,
\end{split}
\label{AnsatzMultipleScale}
\end{equation}
\begin{equation}
\begin{split}
	\partial/\partial  y &\sim \vert \text{M} \vert^{0}
\\
	\partial/\partial  y &\sim \vert \text{M} \vert^{1/2}
\\
	\partial/ \partial z &\sim \vert \text{M} \vert.
\end{split}	
\end{equation} 
Substituting the solution of the type given by Eq.(S\ref{AnsatzMultipleScale}) into the Maxwell equations, Eq.(S\ref{Maxwell3D}), allows to solve each order separately. In particular, in the zeroth order in $\vert \text{M} \vert $ the relevant equations in Eq.(S\ref{AnsatzMultipleScale}) are the first and the third equations, that can be written as
\begin{gather}
 \begin{bmatrix} q_{\alpha}^{2} & i \beta_{0} \frac{\partial}{\partial x} \\ 0 & \frac{\partial^{2}}{\partial x^{2}} - q_{\alpha}^{2} 
 \end{bmatrix} 
 \begin{bmatrix}
& A_{x,\alpha}^{(0)} \\
 & A_{z,\alpha}^{(0)}
 \end{bmatrix} 
 =
 0
\end{gather}
where
\begin{equation}
	q_{\alpha}^{2} = \beta_{0}^{2} - \epsilon_{\alpha}, 
\end{equation}
 and $\alpha = d,m$ .
 The corresponding solution at this order is the textbook SPP solution given by
 \begin{equation}
 \begin{split}
 	A_{x,\alpha}^{(0)} (x,y,z) &= \text{sgn}(x) \frac{i \beta_{0}}{q_{\alpha}} A(y,z) e^{- \text{sgn}(x) q_{\alpha} x}
 \\
 	A_{z,\alpha}^{(0)}(x,y,z) &= A(y,z) e^{- \text{sgn}(x) q_{\alpha} x}
 \end{split}	
 \end{equation}
for both regions $\alpha = d,m$, and where $\text{sgn}(x)$ is the sign function that takes the values $1,-1,0$ for $x>0$, $x<0$ and $x=0$, respectively.

In the order $\vert \text{M} \vert^{1/2}$ the relevant equation is the y-component of the Maxwell equations, $q_{\alpha}^{2}A_{y,\alpha}^{(0)} - \partial^{2}/\partial y^{2} A^{(0)}_{y,\alpha} = 0 $, which yields
\begin{equation}
	A^{(0)}_{y,\alpha} = B(y,z) e^{ - \text{sgn(x)} q_{\alpha} x }
\end{equation}
for both regions $\alpha=d,m$, where $B(y,z)$ is yet to be determined function (note that these solutions satisfies continuity of the tangential component of the electric field across the interface).

In the order $\vert \text{M} \vert$ the governing equations in the metal and the dielectric coincide, respectively, with Eq.(22) and Eq.(24) in \cite{marini2010ginzburg2}, with local nonlinear term, $\chi_{0} \vert A \vert^{2}$ replaced with nonlocal nonlinear term $\text{M} \int d \vec{r}^{\prime}_{\parallel} G_{l}(\vec{r}_{\parallel},\vec{r}_{\parallel}^{\prime}) \vert A (\vec{r}_{\parallel}^{\prime}) \vert^{2}$. Substitution of the corresponding solution into the matching condition yields to cancellation of $B$ and to the following nonlinear and nonlocal Schr{\"o}dinger equation for $A$,
\begin{equation}
	2 i\beta_{0} \dfrac{\partial A}{\partial z} + \dfrac{\partial^{2}A}{\partial y^{2}} + \chi_{TC} A \int d \vec{r}_{\parallel}^{\prime} G_{l} (\vec{r}_{\parallel},\vec{r}_{\parallel}^{\prime}) \vert A(\vec{r}_{\parallel}^{\prime}) \vert^{2} = 0.
\end{equation}


\section{Solution of the local Schr{\"o}dinger equation with harmonic potential}

Assume that the Schr{\"o}dinger equation
\begin{equation}
	2 i \beta_{0} \dfrac{\partial \psi}{\partial z} + \dfrac{\partial^{2} \psi}{\partial y^{2}} + q y^{2} =0
\end{equation}
admits a solution of the Gaussian type, $\psi=b(z)e^{a(z)y^{2}}$. Substitution of the latter into the Schr{\"o}dinger equation leads to a quadratic polynomial in $y^{2}$ and yields
\begin{subequations}
\begin{align}
	a(z) &= \dfrac{1}{2} \sqrt{q} \tan \left( i k_{0} z +  \sqrt{q} c \right) 
\\
	b(z) &= \dfrac{1}{\sqrt{\cos\left( i k_{0} z + \sqrt{q} c \right)}},
\end{align}    
\end{subequations}
where $k_{0}=\sqrt{q}/\beta_{0}$ and $c$ is an integration constant. 
The corresponding squared absolute value of $\psi$ for $c=0$ is given by
\begin{equation}
	\vert \psi \vert^{2} = \dfrac{\sqrt{2}}{f(z)} e^{-\frac{\sinh(2 k^{\prime \prime}_{0} z)}{f(z)}y^{2}}
\label{PsiSquare}    
\end{equation}
where $f(z)$ is a positive function defined as $f(z)=\sqrt{\cosh(2 k_{0}^{\prime} z)+\cos(2 k_{0}^{\prime \prime} z)}$.
As seen from Eq.(\ref{PsiSquare}), the spatial behavior of $\vert \psi \vert^{2}$ along the $y$ axis is determined by the sign of the imaginary part of the constant $k_{0}$. For $q<0$ and $\beta_{0} = \beta_{0}^{\prime} + i \beta_{0}^{\prime \prime}$ with $\beta_{0}^{\prime } > 0$ leads to $k_{0}^{\prime \prime} > 0$ and Gaussian solution. Note that $q<0$ corresponds to $\chi_{TC}<0$, which in turn implies
\begin{equation}
	b \text{M}<0,
\label{bM}	
\end{equation}
 where $\text{M}$ and $b$ are given, respectively, by Eq.(6) and Eq.(9) in the main text. 
The condition, Eq.(S\ref{bM}), splits into the following two non-overlapping conditions:
 
 (i) $\sigma_{T}<0$ and $b>0$, which corresponds to a negative Marangoni constant and the common case $n_{l}>n_{g}$;
 
 (ii) $\sigma_{T}>0$ and $b<0$, which corresponds to a positive Marangoni constant and a case in which the gas is replaced by a liquid dielectric
 with larger index of refraction than $n_{l}$.

\section{Numerical simulation details for Fig.3}

Under the scaling to dimensionless variables $\vec{r}_{\parallel} \rightarrow L \vec{r}_{\parallel}$, $t \rightarrow (L^{4}/D_{l} \ell_{c}^{2}) t$, $\eta \rightarrow h_{0} \eta$ and $I \rightarrow I_{0} I$, Eq.(S\ref{ThinFilmEqU3}) takes the form 
\begin{equation}
	\dfrac{\partial ( \eta/h_{0})}{\partial t} + \nabla^{4} (\eta/h_{0}) = q I 
\end{equation}
where $q = - 3 L^{4}/(2 h_{0}^{2}) (\sigma_{T}/\sigma_{0}) (\alpha_{th}^{m}/k_{th}^{m}) I_{0}$ 
Eq.(S\ref{GreenActuation})
Here, we have assumed that the typical scale of the fluid deformation, $L$,  satisfies $\ell_{c}/L \ll 1$, which allows to drop the second order derivative term in Eq.(\ref{ThinFilmEqU33}). 

\begin{equation}
\begin{split}
	& \Delta \epsilon_{D} = b \eta/h_{0};
 \\   
  & b = 4q_{D}h_{0} (n_{l}-n_{g}) \Big[ n_{l} - (n_{l}-n_{g})e^{-2q_{D}h_{0}} \Big] e^{-2q_{D}h_{0}}
\end{split}    
\end{equation}
For $n_{l}=2$, $n_{g}=1$, $h_{0}=200$ nm, $\lambda=800$ nm, and dielectric constant of gold at that wavelength $\epsilon=-24 + 1.5 i$, and $b=1$. 
Fig.(3) presents simulation results of two SPP's with initially Gaussian lateral intensity distribution, and $\eta/h_{0}=0.2$ which attained at dimensional time $t=0.1 \cdot L^{4}/(D_{l} \ell_{c}^{2})$.  
The fluid deformation is obtained by convolving Gaussian intensity distribution with the the time integrated Green's function, given by Eq.(S\ref{GreenActuation}). Numerical integration is accomplished by utilizing numerical solver \textit{Mathematica v.11} and implementing its built-in Explicit Runge Kutta method. The size of the simulation domain along the $y$ direction is $60$ (in dimensionless units normalized by $k_{0}$) and the SPP envelope, $A$, is subject to a vanishing boundary conditions on the lines $y=\pm 30$. The spacing of the numerical grid is $0.1$ and the number of effective digits of precision is three.

%
 

\end{document}